\documentclass[prb,twocolumn,showpacs,superscriptaddress,groupedaddress,amsmath,amssymb]{revtex4-1}

\usepackage{graphicx}
\usepackage{dcolumn}
\usepackage{bm}

\usepackage{hyperref}
\hypersetup{colorlinks = true, 
  urlcolor = blue, linkcolor = blue, citecolor = blue}

\usepackage{color}
\usepackage{ulem} 

\begin{document}

\preprint{APS/123-QED}

\title{
Multiband Metallic Ground State in Multilayered Nickelates La$_3$Ni$_2$O$_7$ and La$_4$Ni$_3$O$_{10}$ \\ Probed by $^{139}$La-NMR at Ambient Pressure
}

\author{Masataka Kakoi$^{1,2}$, Takashi Oi$^2$, Yujiro Ohshita$^2$, Mitsuharu Yashima$^2$, \\Kazuhiko Kuroki$^1$, Takeru Kato$^2$, Hidefumi Takahashi$^2$, Shintaro Ishiwata$^2$, \\Yoshinobu Adachi$^3$, Naoyuki Hatada$^3$, Tetsuya Uda$^3$, Hidekazu Mukuda$^2$}

\affiliation{$^1$Department of Physics, Osaka University, Toyonaka, Osaka 560-0043, Japan\\
$^2$Graduate School of Engineering Science, Osaka University, Toyonaka, Osaka 560-8531, Japan\\
$^3$Department of Materials Science and Engineering, Kyoto University, Kyoto, 606-8501, Japan
}
\date{\today}

\begin{abstract}
We report a $^{139}$La-NMR study of polycrystalline samples of multi($n$)-layered nickelates, La$_3$Ni$_2$O$_{7-\delta}$ ($n=2$) and La$_4$Ni$_3$O$_{10-\delta}$ ($n=3$), at ambient pressure.
Measurements of the nuclear magnetic resonance (NMR) spectra and nuclear spin relaxation rate ($1/T_1$) indicate the emergence of a density wave order with a gap below $T^*\sim150$~K for La$_3$Ni$_2$O$_{7-\delta}$ and $\sim130$~K for La$_4$Ni$_3$O$_{10-\delta}$. 
The finite value of $1/T_1$ below $T^*$ indicates metallic ground states with the remaining density of states at the Fermi level ($E_{\rm F}$) under the density wave order. These features are attributed to multiple $d$ electron bands with different characteristics. 
Above $T^*$, the gradual decrease in $1/T_1T$ upon cooling implies the presence of a band with flat dispersion near $E_{\rm F}$.
From our microscopic probes, we point out that these nickelates ($n=2$ and $3$) possess similar electronic states despite the difference in the formal valence of the Ni $d$ electron states, which provides a basis for understanding the novel high-$T_{\rm c}$ superconductivity under high pressures. 
\end{abstract}

\maketitle

After the discovery of high transition temperature (high-$T_{\rm c}$) superconductivity in cuprates, novel high-$T_{\rm c}$ states have been explored in other transition metal oxides.
The parent phase of cuprates is characterized by a Cu-$3d^9$ electron configuration, which results in an antiferromagnetic insulator owing to the strong electron correlations in the $d_{x^2-y^2}$ orbital band. 
The slight carrier doping in the CuO$_2$ plane, which suppressed the antiferromagnetic phase, led to the emergence of a high-$T_{\rm c}$ state. 
Regarding the mechanism of the high $T_{\rm c}$ cuprates, a strong electron correlation within the $d_{x^2-y^2}$ orbital in a nearly $d^9$ electron configuration is considered indispensable.
Since then, the realization of the $d^9$ state had been anticipated in Ni oxides. However, the Ni$^{1+}$($d^9$) state tends to be unstable compared to the Ni$^{2+}$($d^8$) or Ni$^{3+}$($d^7$) states. 
In 2019, superconductivity in nickelate was discovered in a hole-doped infinite-layer NdNiO$_2$ close to the Ni-$d^9$ electron configuration.~\cite{D-Li_2019}
Nevertheless, the maximum $T_{\rm c}$ was still much lower than that of most cuprates.

Recently, remarkably high-$T_{\rm c}$ superconductivity with a maximum $T_{\rm c}$ of approximately $80$~K has been reported for layered nickelate La$_3$Ni$_2$O$_7$ under high pressure of $\sim15$~GPa,~\cite{Sun_2023} and it has also been confirmed by other groups.~\cite{Hou_replication_2023,Yanan-Zhang_replication_preprint,Sakakibara_La4310_2024,G-Wang_2024,Wang_La2PrNi2O7_preprint,Y-Zhou_replication_preprint,Puphal_replication_preprint}
This compound belongs to the Ruddlesden-Popper (RP) phase, La$_{n+1}$Ni$_{n}$O$_{3n+1}$ composed of $n$-layered NiO$_6$ octahedra.
Compounds with $n\ge2$ have an intermediate electron configuration between Ni-$d^8$ and Ni-$d^7$ in the formal valence, with partially occupied electrons in the $d_{x^2-y^2}$ and $d_{3z^2-r^2}$ orbitals. 
Multiple degrees of freedom in $d$ electrons, such as charge, spin, and orbital, are expected to play a role in the high-$T_{\rm c}$ phase of La$_3$Ni$_2$O$_7$.
Theoretically, the electronic state of La$_3$Ni$_2$O$_7$ is considered to be described by a bilayer model with a large interlayer hopping between the $d_{3z^2-r^2}$ orbitals coupled with the $d_{x^2-y^2}$ orbital.
However, the interplay between these two orbitals remains controversial.~\cite{Nakata_2017,Luo_2023,Y-Zhang_2023,QG-Yang_2023,Sakakibara_flex_2024,YB-Liu_2023, W-Wu_2024, C-Lu_theory_preprint, XZ-Qu_2024}
The rich variation in elements and layer numbers ($n$) within the RP series would offer high potential for novel high-$T_{\rm c}$ research.
Recently, it was reported that $n=3$ in the RP phase La$_4$Ni$_3$O$_{10}$, shows signatures of superconductivity at approximately $25$~K under high pressures.~\cite{Sakakibara_La4310_2024, Q-Li_4310_2024, Y-Zhu_4310_preprint, M-Zhang_4310_preprint}
Conducting systematic experiments on the $n$-layered RP phases is crucial for understanding the parent electronic states and origins of high-$T_{\rm c}$ superconductivity in nickelates under pressure.

In this Letter, we report the $^{139}$La-nuclear magnetic resonance (NMR) study of $n$-layered nickelates La$_3$Ni$_2$O$_{7-\delta}$ ($n=2$) and La$_4$Ni$_3$O$_{10-\delta}$ ($n=3$) at ambient pressure. 
The measurements of the NMR spectra and $1/T_1$ indicate the feature of the multiple bands: one exhibited a density wave order below $T^*\sim150$~K for La$_3$Ni$_2$O$_{7-\delta}$ and $\sim130$~K for La$_4$Ni$_3$O$_{10-\delta}$, while the other retained its metallic state even below $T^*$. 
These results indicate a qualitative resemblance between the electronic states of the $n$-layered nickelates in the RP phase ($n=2$ and $3$) at ambient pressure. 

Polycrystalline samples of La$_3$Ni$_2$O$_{7-\delta}$ and La$_4$Ni$_3$O$_{10-\delta}$ were synthesized using the fine-grained precursors La$_2$NiO$_4$ and NiO, along with post-sintering oxidation processes, as described previously.~\cite{Adachi_2019}
The oxygen deficiencies were estimated from thermogravimetric analysis to be $\delta\simeq0.03$ for La$_3$Ni$_2$O$_{7-\delta}$ and $\delta\simeq0.36$ for La$_4$Ni$_3$O$_{10-\delta}$.
The bulk resistivity of both the samples is provided in the Supplemental Material\cite{SM}.
The $^{139}$La($I=7/2$)NMR measurements were performed on coarse powder samples. 
$^{139}$La-NMR spectra were obtained by sweeping the field at a fixed frequency of $f_0 = 45$~MHz.
The nuclear spin-lattice relaxation rate ($1/T_1$) was measured at central transition ($+1/2 \leftrightarrow -1/2$).
This was determined by fitting the recovery curve for $^{139}$La nuclear magnetization to a multiple exponential function:
\begin{align}\label{eq:recovery-curve}
    \nonumber
       1-M(t)/M(\infty)  &= c_1\exp(-t/T_1) + c_2\exp(-6t/T_1) \\
     \nonumber
               &\quad + c_3\exp(-15t/T_1) + c_4\exp(-28t/T_1),
\end{align}
where $M(t)$ is the nuclear magnetization after time $t$ from the NMR saturation pulse, and the coefficients $c_1=1/84$, $c_2=3/44$, $c_3=75/364$, and $c_4=1225/1716$.~\cite{Narath_1967}

\begin{figure}[tbp]
  \includegraphics[width=\linewidth]{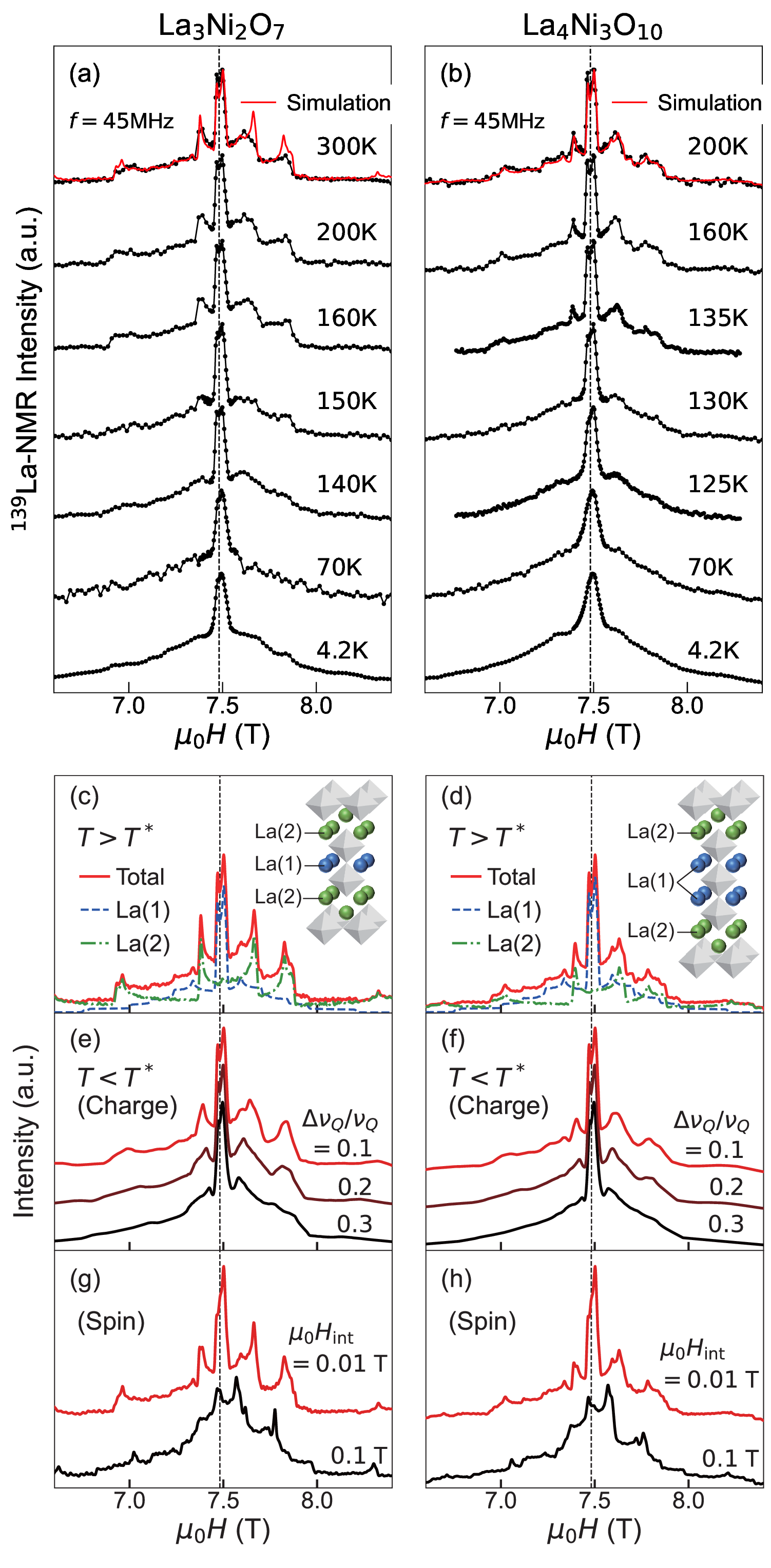}
  \caption{\label{fig:NMR-spectra}
  (Color online)
  $T$ dependence of $^{139}$La-NMR spectra for (a)~La327 and (b)~La4310.
  The red curves are simulations of the spectra at high temperatures, which are well reproduced by the superposition of the spectra for two inequivalent La($i$) sites (see (c)~La327 and (d)~La4310) with the appropriate values of $\nu_Q(i)$ and $\eta(i)$ (see Table~\ref{tab:simulation_parameters}). 
  (e)-(f) Broad spectra below $T^*$ is simulated by assuming the uniform distribution of $\nu_Q(i)$ in the range of [$\nu_Q(i) - \Delta \nu_Q(i)$,\, $\nu_Q(i) + \Delta \nu_Q(i)$] for (e)~La327 and (f)~La4310.
  (g)-(h) Simulations assuming the internal fields perpendicular to $c$-axis $\mu_0H_{\rm int(La)}=\pm0.01$~T and $\pm0.1$~T at both La($i$) sites, which are inconsistent with the spectra below $T^*$.
  }
\end{figure}

Figures~\ref{fig:NMR-spectra}(a) and \ref{fig:NMR-spectra}(b) show the temperature ($T$) dependence of the $^{139}$La-NMR spectra for La$_3$Ni$_2$O$_{7-\delta}$~(La327) and La$_4$Ni$_3$O$_{10-\delta}$~(La4310), respectively. 
The spectra show typical powder patterns at temperatures above $T^*\sim150$~K and $\sim130$~K for La327 and La4310, respectively, articulated by nuclear quadrupole interactions. 
There are two crystallographically-inequivalent La sites located at the inner (La(1)) and outer (La(2)) positions of $n$-layered NiO$_6$ perovskite blocks (see the insets of Figs.~\ref{fig:NMR-spectra}(c) and \ref{fig:NMR-spectra}(d)).
As shown in Figs.~\ref{fig:NMR-spectra}(c) and \ref{fig:NMR-spectra}(d), the observed NMR spectra at high temperatures were well reproduced by the superposition of the spectra for two La($i$) sites with the values of $\nu_Q(i)$ and $\eta(i)$ given in Table~\ref{tab:simulation_parameters}.
Below, $\nu_Q(i)$ are the nuclear quadrupole resonance (NQR) frequencies defined by $\nu_Q(i) = 3eV_{zz}(i)Q/(2I(2I-1)h)$ for each La($i$) site, where $V_{\mu\mu}(i)$ is the principal value of the electric field gradient tensor for the La($i$) site; $Q$ is the electric quadrupole moment of the $^{139}$La nucleus and the asymmetry parameter is $\eta(i)=|V_{xx}-V_{yy}|/V_{zz}$. 
The intensity ratios observed at the La(1) and La(2) sites in the superimposed spectra shown in Fig.~\ref{fig:NMR-spectra}(c) and \ref{fig:NMR-spectra}(d) are approximately equivalent to the ratio of the number of La sites in La327 and La4310, which is, $1:2$ for La327 and $1:1$ for La4310.
The observed value of $\nu_Q$ at La(1) indicates a higher symmetry around the La(1) site in comparison to the La(2) site.
The finite value of $\eta$ indicates the orthorhombicity of these compounds at ambient pressure, which should be zero in the case of tetragonal symmetry. 
The Knight shifts for both La sites in both samples were $K\sim0.0\%$, and their $T$-dependence was negligibly small at this experimental resolution.  
The obtained values $\nu_Q(i)$, $\eta(i)$, and $K$ were similar to those reported previously.~\cite{Fukamachi_2001}
\begin{table}[b]
    \caption{
    The NQR frequencies $\nu_Q(i)$ and asymmetry parameters $\eta(i)$ for two inequivalent La($i$) sites, namely, La327 and La4310 obtained from the spectrum analyses.
    \label{tab:simulation_parameters}}
    \begin{ruledtabular}
    \renewcommand{\arraystretch}{1.2}
    \begin{tabular}{lcccc}
    & \multicolumn{2}{c}{La(1)} & \multicolumn{2}{c}{La(2)} \\
    & $\nu_Q(1)$ [MHz] & $\eta(1)$ & $\nu_Q(2)$ [MHz]& $\eta(2)$ \\
    \hline
    La$_3$Ni$_2$O$_{7-\delta}$ & $2.2$ & $0.30$ & $5.6$ & $0.05$\\
    La$_4$Ni$_3$O$_{10-\delta}$ & $2.2$ & $0.30$ & $5.2$ & $0.10$ \\
    \end{tabular}
    \end{ruledtabular}
\end{table}

Then, we focused on the broad spectra at the temperatures below $T^*\sim150$~K for La327 and $\sim130$~K for La4310 (see Figs.~\ref{fig:NMR-spectra}(a) and \ref{fig:NMR-spectra}(b)).
As indicated in many studies, this broadening originates from the possible orders of charge and/or spin.~\cite{Seo_1996,Carvalho_2000,GQ-Wu_2001,HX-Li_ARPES_2017,Kumar_2020,J-Zhang_2020_nat-commun,J-Zhang_2020-PRMat,N-Yuan_2024,ZJ-Liu_2023,Liu_optical_preprint,T-Xie_Neutron_preprint,K-Chen_muSR_preprint,X-Chen_Xray_preprint,Z-Dan_NMR_preprint,Khasanov_muSR_preprint}
In these nickelates, a complex state of charge and spin degrees of freedom may be inevitable because of the partial occupation of multiple $d$-electron bands.
First, we assume the simplest case, where all broadening is attributed to the possible charge density wave (CDW) order at the Ni sites. 
Local charge inhomogeneity results from the $\nu_Q(i)$ distribution at both La sites because it is proportional to the electric field gradient.
Figures~\ref{fig:NMR-spectra}(e) and \ref{fig:NMR-spectra}(f) show simulations assuming a distribution of $\nu_Q(i)$ in the range of [$\nu_Q(i)-\Delta\nu_Q(i),\, \nu_Q(i)+\Delta\nu_Q(i)$] for two La($i$) sites. 
The feature of the simulations is consistent with the typical experimental results, where broadening is more significant at the satellite peaks than at the center peak. 
Assuming that $\Delta \nu_Q(i)/\nu_Q(i) = 0.3$ is similar to that of the observed spectra.
Next, we consider another case in which all the broadening is attributed to a possible spin density wave (SDW) order.
Figures~\ref{fig:NMR-spectra}(g) and \ref{fig:NMR-spectra}(h) show the simulations assuming internal fields $\mu_0H_{\rm int(La)}$ of $\pm0.01$~T and $\pm0.1$~T, respectively, at both La($i$) sites for La327(La4310), for example.
The overall spectra at low temperatures could not be reproduced in the presence of an internal field alone. 
Note that we do not rule out the presence of an SDW order with a small internal field ($<0.01$~T) under the charge order.
The possibility of the SDW order coexisting with the charge order~\cite{J-Zhang_2020_nat-commun, ZJ-Liu_2023} is discussed again in the final paragraph.

\begin{figure}[t]
\includegraphics[width=\linewidth]{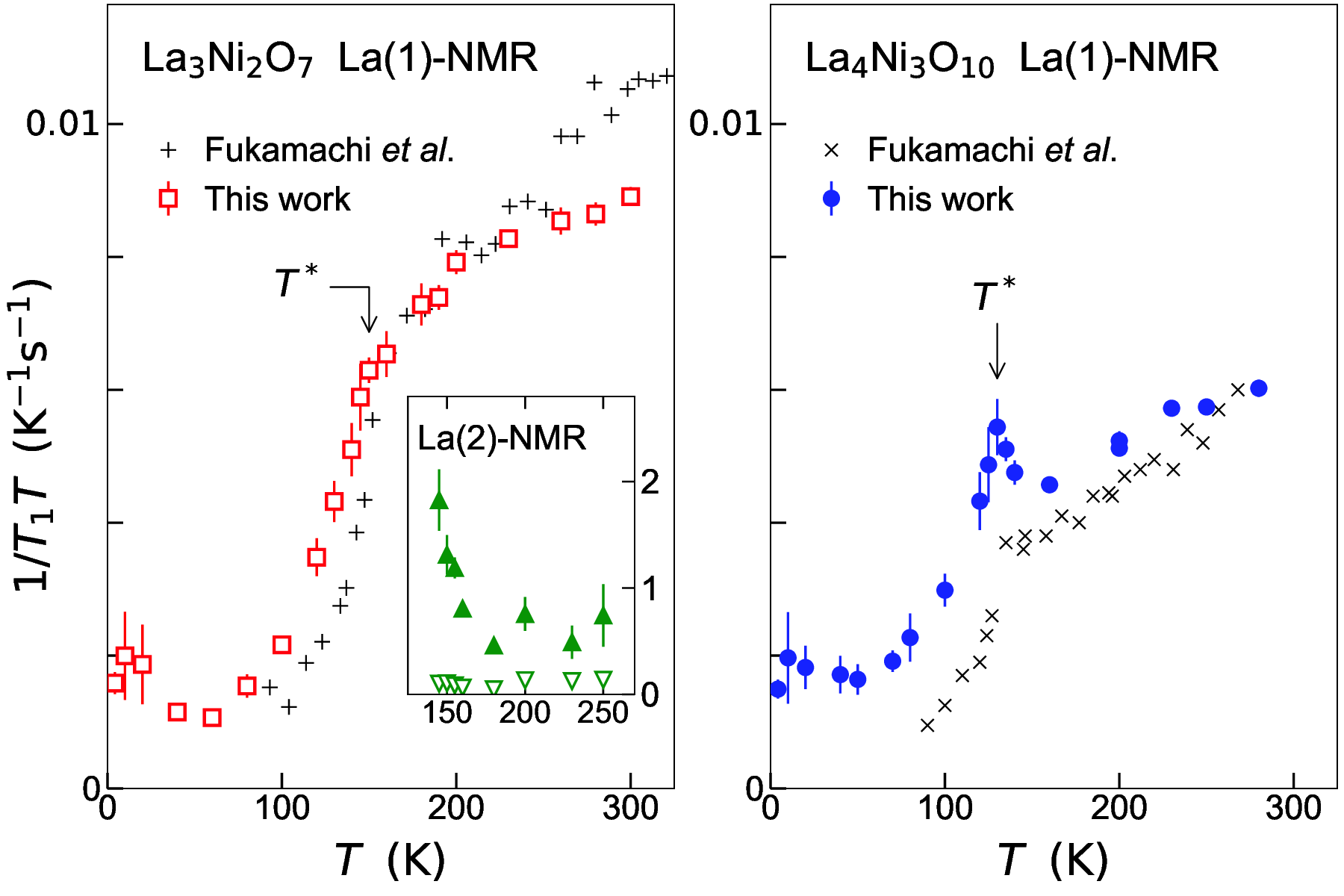}
\caption{\label{fig:T1}
(Color online)
$T$ dependence of $1/T_1T$ measured at the La(1) site for La327 and La4310.
Crosses denote the previous data reported by Fukamachi {\it et al}.~\cite{Fukamachi_2001}
The inset shows the $1/T_1T$ measured at La(2) site in La327, which is dominated by two $T_1$ components, $T_{1S}$($\blacktriangle$) and $T_{1L}$($\triangledown$).
}
\end{figure}

\begin{figure}[t]
  \includegraphics[width=0.85\linewidth]{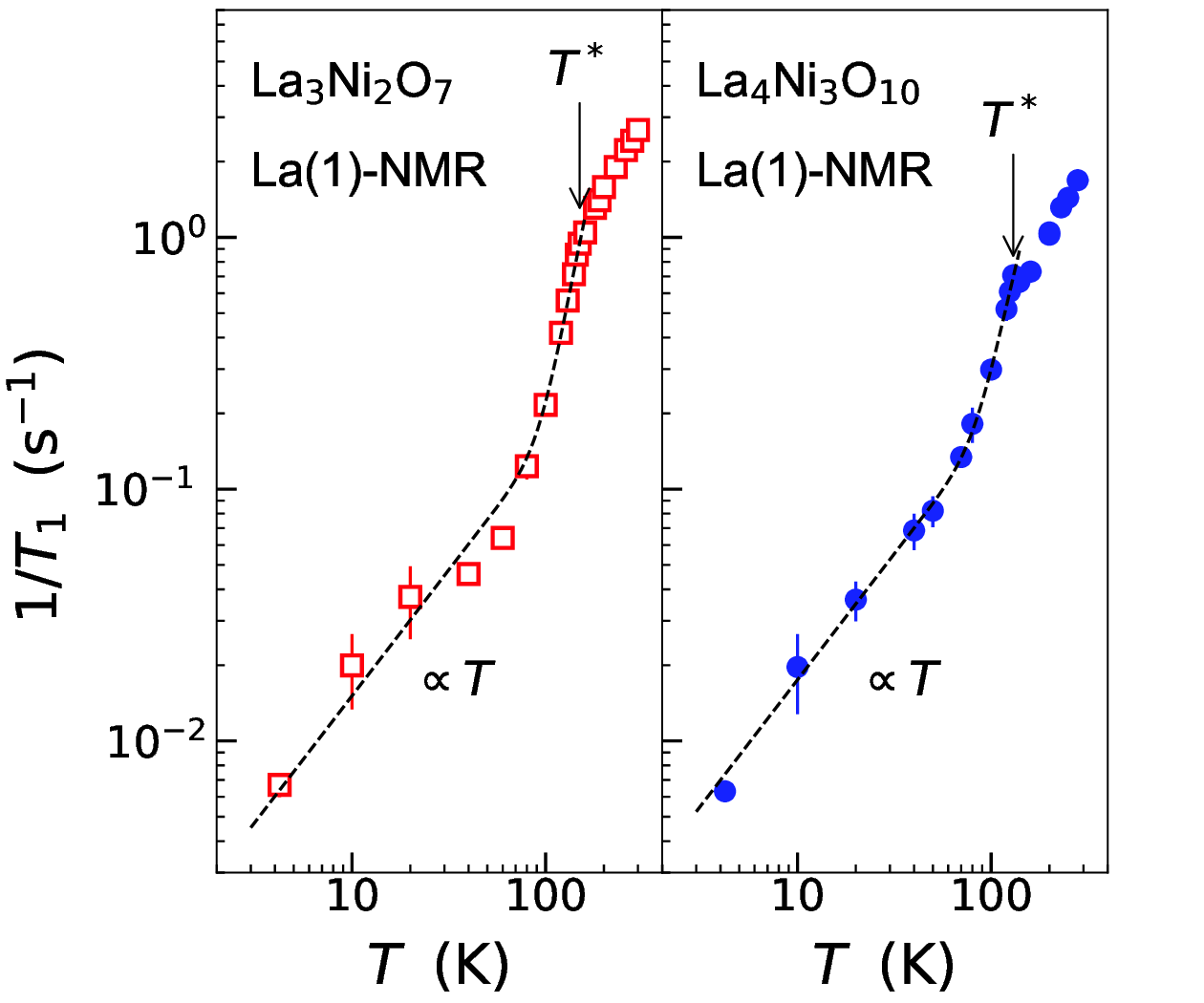}
  \caption{\label{fig:T1-fit}
  (Color online)
  $T$ dependence of $1/T_1$ below $T^*$ is reproduced by assuming Eq.~(\ref{eq:partial_gap}) with the parameters of $a\sim1.5(1.7)\times10^{-3}$ and $\Delta\sim60(50)$~meV for La327(La4310) (see the dashed lines). 
  }
\end{figure}

Next, we discuss the nuclear spin-relaxation rates $1/T_1$ measured at the La(1) site. 
As shown in Fig.~\ref{fig:T1}, the $T$ dependence of $1/T_1T$ exhibits a distinct anomaly at $T^*$, which coincides with the appearance of broad spectra.
The sharp drop in $1/T_1T$ below $T^*$ indicates the onset of a gap at $E_{\rm F}$, owing to the possible density wave order. 
At low temperatures ($T\ll T^*$), $1/T_1T$ remains finite in both samples, indicating that a finite density of state (DOS) at $E_{\rm F}$ remains even under the density wave order.
Here, we attempt to fit our experimental results at low temperatures by assuming the following relation:
\begin{equation}\label{eq:partial_gap}
    \frac{1}{T_1} \simeq aT + b\exp\left(-\frac{\Delta}{k_BT}\right),
\end{equation}
The first term corresponds to the metallic bands remaining even below $T^*$ while the latter corresponds to the bands with a gap of $\Delta$ at $E_{\rm F}$.
As shown in Fig.~\ref{fig:T1-fit}, the experimental data below $T^*$ were reproduced for gap sizes of $\Delta\sim60$~meV and $\sim50$~meV for La327 and La4310, respectively.  
Similar values have been previously reported for the optical conductivity of La327 ($\Delta\sim100$~meV)~\cite{Liu_optical_preprint} and ARPES for La4310 ($\Delta\sim40$~meV).~\cite{HX-Li_ARPES_2017}
According to these experimental studies, combined with band calculations, it was reported that the bands near $E_{\rm F}$ are composed of two characteristic Ni-$3d$ orbitals: one is a bonding band with a flat dispersion derived from the $d_{3z^2-r^2}$ orbital and the other is a band with broad dispersion derived from the $d_{x^2-y^2}$ orbital for both La327 and La4310.~\cite{Yang_ARPES_preprint,Liu_optical_preprint,HX-Li_ARPES_2017}
The former could give rise to a strong variation in the DOS near $E_{\rm F}$, owing to flat dispersion.~\cite{Jung_2022,Sun_2023}
When a DOS with such characteristics is located slightly away from $E_{\rm F}$, we expect a gradual decrease in $1/T_1T$ upon cooling.~\cite{Mukuda_2014}
As shown in Fig.~\ref{fig:T1}, this can be observed in the $T$ range of $T>T^*$ in both compounds. 
On the other hand, the latter would yield normal metallic excitations at $E_{\rm F}$, which is characterized by a finite value of $1/T_1T$.
This behavior is experimentally observed below $T^*$ for both compounds. 
These two distinct features can be attributed to the unique multiband characteristics of La327 and La4310.

Finally, we note that some studies pointed to the possible SDW order for La4310~\cite{J-Zhang_2020_nat-commun} and La327.~\cite{ZJ-Liu_2023,K-Chen_muSR_preprint,X-Chen_Xray_preprint,Z-Dan_NMR_preprint,Khasanov_muSR_preprint}
In the case of La$_2$NiO$_4$ (La214) ($n=1$), the coexistence of the spin and charge orders has been reported.~\cite{Kajimoto_2003}
The $^{139}$La-NMR spectra of La214 are significantly affected by the large internal field ($\sim 2$~T) from the magnetic moment at the Ni sites, owing to the large hyperfine coupling constant through the strong covalency of Ni($3d_{3z^2-r^2}$)- apical O($2p_z$)-La($6s$).~\cite{Wada_1993}
The La214 possesses a single La site corresponding to the La(2) site in the notation used in this paper.
The inset of Fig.~\ref{fig:T1} shows the preliminary results of $1/T_1T$ measured at the La(2) site, which is two orders of magnitude larger than that at the La(1) site owing to the large fluctuating hyperfine field from the spins at the Ni sites.
Here, we observe shorter and longer $T_1$ components denoted by $T_{1S}$ and $T_{1L}$, respectively.
$1/T_{1S}$ was critically enhanced towards $T^*$, suggesting the possibility of magnetic fluctuations developing just above $T^*$.
$1/T_{1L}$ also remained large, in contrast to that of the La(1) site, suggesting that the La(2) site may be a good probe for the magnetic properties of these nickelates.
In contrast, the La(1) site is likely insensitive to magnetic fluctuations because the eight neighboring Ni sites mostly cancel the magnetic hyperfine field.
Although the broadening of the spectra below $T^*$ is primarily attributed to charge anomalies (see Fig.~\ref{fig:NMR-spectra}), positing distinct internal fields at the two La sites, namely, a small field at the La(1) site and a larger field at the La(2) site, might be another possible scenario for reproducing the spectra.
To identify the further details of the possible charge and/or spin orders, La-NQR measurements, which distinguish La(1) and La(2) at zero field, will be necessary in future studies.
Note that a very small internal field due to the SDW order has recently been reported by NMR in single crystals of La327.~\cite{Z-Dan_NMR_preprint} 
There remains the problem of sample dependence, such as the local electronic states of Ni around the oxygen-deficient sites.  
Its relationship with the local microscopic states around oxygen-deficient sites should clarify this in the future.

In summary, our $^{139}$La-NMR study of polycrystalline samples of $n$-layered nickelates La$_{n+1}$Ni$_{n}$O$_{3n+1}$ ($n=2$ and $3$) reveals that the density wave order associated with the charges emerges below $T^*\sim150(130)$~K for La327(La4310), based on the measurements of the NMR spectra and $1/T_1$. 
The behavior of the finite value of $1/T_1T$ indicates a metallic ground state with a finite DOS at $E_{\rm F}$ even below $T^*$. 
Considering the band calculations, this can be attributed to multiple bands, that is, the metallic state dominated by the $d_{x^2-y^2}$ orbital and the density wave state dominated by the $d_{3z^2-r^2}$ orbital. 
In particular, the presence of a strong variation in the DOS near $E_{\rm F}$ is suggested by the gradual decrease in $1/T_1T$ upon cooling above $T^*$, which can be attributed to the $d_{3z^2-r^2}$ band with a flat dispersion near $E_{\rm F}$.
Our microscopic NMR probes demonstrate that these nickelates ($n=2$ and $3$) possess similar electronic states despite the difference in the electronic configuration of the Ni-$3d$ electrons in the formal valence state. 
The universality and diversity of $n$-layered nickelates of the RP phase might be crucial for understanding the novel high-$T_{\rm c}$ superconductivity at high pressures. 

{\footnotesize 
    This work was partly supported by Izumi science and technology foundation, Casio science promotion foundation, and Takahashi Industrial and Economic Research Foundation.
    One of the authors (M.K.) was supported by Program for Leading Graduate Schools: "Interactive Materials Science Cadet Program".
}

\bibliography{18019Refs}

\clearpage
\onecolumngrid

\section*{Supplemental Material}

\subsection*{Resistivity in our polycrystalline samples of La$_3$Ni$_2$O$_{7-\delta}$ and La$_4$Ni$_3$O$_{10-\delta}$}

\begin{figure}[htbp]
  \renewcommand{\thefigure}{S1}
  \includegraphics[width=0.65\linewidth]{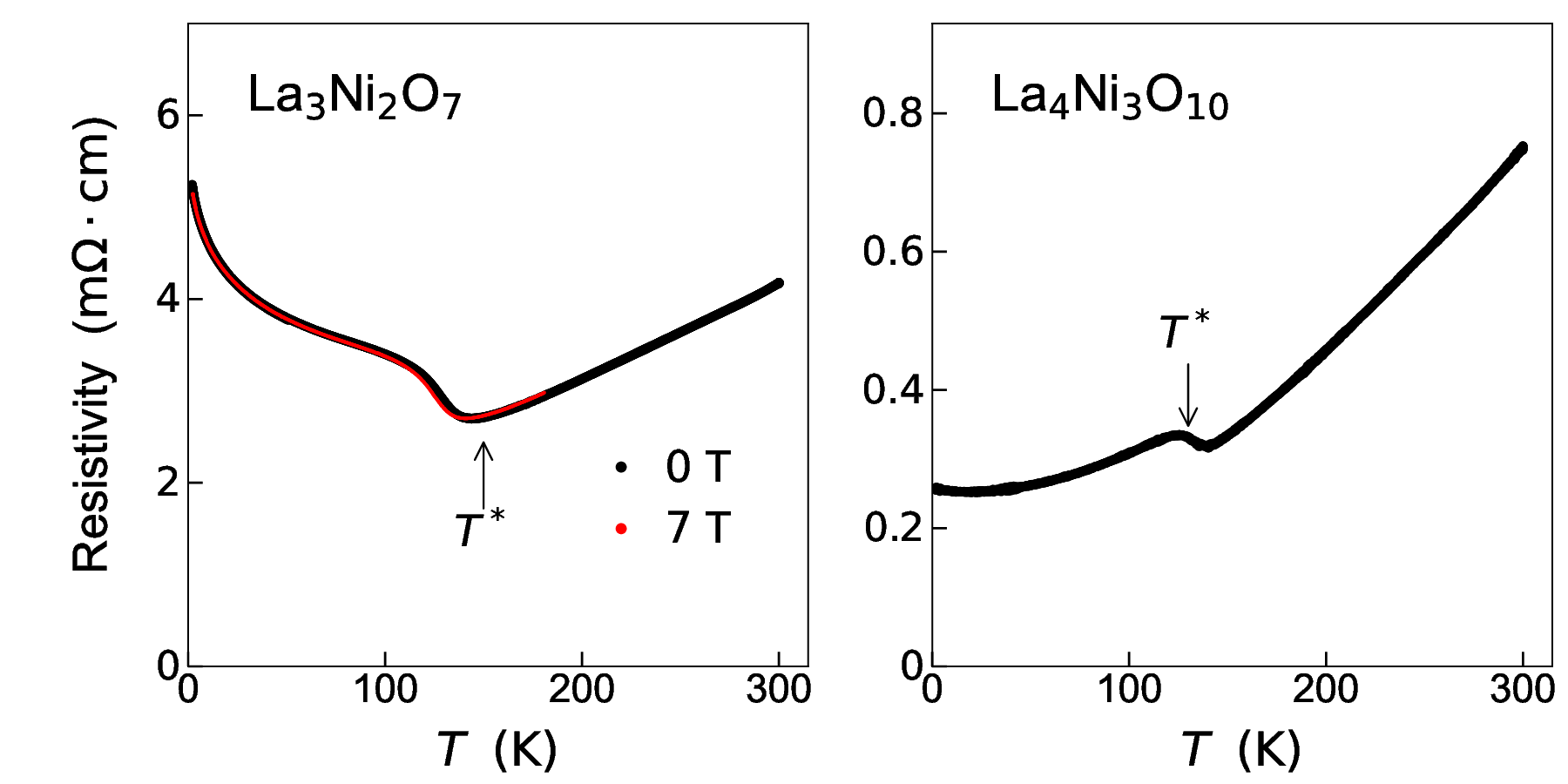}
  \caption{\label{fig:resistivity}
  Electrical resistivity for (left) La327 and (right) La4310. 
  The arrow at $T^*=150(130)$~K for La327(La4310) indicates the density wave transition temperature detected by the $^{139}$La-NMR measurement.
  For La327, the resistivity data under external fields of $7$~T is shown.
  }
\end{figure}

Figure~\ref{fig:resistivity} shows the resistivity in polycrystalline samples of La$_3$Ni$_2$O$_{7-\delta}$~(La327) and La$_4$Ni$_3$O$_{10-\delta}$~(La4310) measured at ambient pressure.
The anomalies are clearly observed at the density wave transition temperature $T^*$ for both samples.
The resistivity of La327 shows a small upturn below $T^*$, but its value of resistivity is kept within the order of several ${\rm m\Omega\cdot cm}$.
These facts are consistent with the multiband picture with the metallic state under the density wave order in both compounds.
The resistivity in La327 is insensitive to external fields up to $7$~T.

\end{document}